\documentclass[aps,prl,twocolumn,superscriptaddress,longbibliography,bibnotes]{revtex4-2}
\usepackage{diagbox}
\usepackage{amsmath}
\usepackage{amsfonts}
\usepackage{physics}
\usepackage[space]{grffile}
\usepackage{graphicx} % Include figure files
\usepackage{amssymb}
\usepackage{upgreek}
\usepackage{lineno}
\usepackage{natbib}
\usepackage{braket}
\usepackage{float}
\usepackage{color}
\usepackage{blindtext}
\usepackage{comment}
\usepackage{graphicx}
\usepackage{xr-hyper}
\usepackage{hyperref}
\usepackage{xcolor}
\hypersetup{colorlinks,breaklinks,
            urlcolor=[rgb]{0,0,0.64},
            linkcolor=[rgb]{0,0,0.64},
            citecolor=[rgb]{0,0,0.64},
            filecolor=[rgb]{0,0,0.64}}
\usepackage[T1]{fontenc}
\usepackage [latin1]{inputenc}
 

\begin{document}
\title{Bidirectional ultrafast control of charge density waves via phase competition}

\author{Honglie Ning}
\thanks{These authors contributed equally to this work}
\affiliation{Department of Physics, Massachusetts Institute of Technology, Cambridge, MA 02139, USA}

\author{Kyoung Hun Oh}
\thanks{These authors contributed equally to this work}
\affiliation{Department of Physics, Massachusetts Institute of Technology, Cambridge, MA 02139, USA}

\author{Yifan Su}
\thanks{These authors contributed equally to this work}
\affiliation{Department of Physics, Massachusetts Institute of Technology, Cambridge, MA 02139, USA}

\author{Zhengyan Darius Shi}
\affiliation{Department of Physics, Massachusetts Institute of Technology, Cambridge, MA 02139, USA}

\author{Dong Wu}
\affiliation{Beijing Academy of Quantum Information Sciences, Beijing 100913, China}

\author{Qiaomei Liu}
\affiliation{International Center for Quantum Materials, School of Physics, Peking University, Beijing 100871, China}

\author{B. Q. Lv}
\affiliation{Department of Physics, Massachusetts Institute of Technology, Cambridge, MA 02139, USA}
\affiliation{Tsung-Dao Lee Institute, School of Physics and Astronomy, and Zhangjiang Institute for Advanced Study, Shanghai Jiao Tong University, Shanghai 200240, China}

\author{Alfred Zong}
\affiliation{Department of Physics, Massachusetts Institute of Technology, Cambridge, MA 02139, USA}
\affiliation{Departments of Physics and of Applied Physics, Stanford University, Stanford, CA 94305, USA}
\affiliation{Stanford Institute for Materials and Energy Sciences, SLAC National Accelerator Laboratory, Menlo Park, CA 94025, USA}

\author{Gyeongbo Kang}
\affiliation{PAL-XFEL, Pohang Accelerator Laboratory, Pohang, Gyeongbuk 37673, Republic of Korea}

\author{Hyeongi Choi}
\affiliation{PAL-XFEL, Pohang Accelerator Laboratory, Pohang, Gyeongbuk 37673, Republic of Korea}

\author{Hyun-Woo J. Kim}
\affiliation{Department of Physics, Pohang University of Science and Technology, Pohang 37673, Republic of Korea}

\author{Seunghyeok Ha}
\affiliation{Department of Physics, Pohang University of Science and Technology, Pohang 37673, Republic of Korea}

\author{Jaehwon Kim}
\affiliation{Department of Physics, Pohang University of Science and Technology, Pohang 37673, Republic of Korea}

\author{Suchismita Sarker}
\affiliation{CHESS, Cornell University, Ithaca, New York 14853, USA}

\author{Jacob P. C. Ruff}
\affiliation{CHESS, Cornell University, Ithaca, New York 14853, USA}

\author{B. J. Kim}
\affiliation{Department of Physics, Pohang University of Science and Technology, Pohang 37673, Republic of Korea}

\author{N. L. Wang}
\affiliation{Beijing Academy of Quantum Information Sciences, Beijing 100913, China}
\affiliation{International Center for Quantum Materials, School of Physics, Peking University, Beijing 100871, China}

\author{Todadri Senthil}
\affiliation{Department of Physics, Massachusetts Institute of Technology, Cambridge, MA 02139, USA}

\author{Hoyoung Jang}
\affiliation{PAL-XFEL, Pohang Accelerator Laboratory, Pohang, Gyeongbuk 37673, Republic of Korea}

\author{Nuh Gedik}
\email[email: ]{gedik@mit.edu}
\affiliation{Department of Physics, Massachusetts Institute of Technology, Cambridge, MA 02139, USA}

\begin{abstract}
The intricate competition between coexisting charge density waves (CDWs) can lead to rich phenomena, offering unique opportunities for phase manipulation through electromagnetic stimuli. Leveraging time-resolved X-ray diffraction, we demonstrate ultrafast control of a CDW in EuTe$_4$ upon optical excitation. At low excitation intensities, the amplitude of one of the coexisting CDW orders increases at the expense of the competing CDW, whereas at high intensities, it exhibits a nonmonotonic temporal evolution characterized by both enhancement and reduction. This transient bidirectional controllability, tunable by adjusting photo-excitation intensity, arises from the interplay between optical quenching and phase-competition-induced enhancement. Our findings, supported by phenomenological time-dependent Ginzburg-Landau theory simulations, not only clarify the relationship between the two CDWs in EuTe$_4$, but also highlight the versatility of optical control over order parameters enabled by phase competition.
\end{abstract}

\maketitle

Understanding the interplay between neighboring phases is a cornerstone of condensed matter physics, providing pathways for on-demand phase control through external perturbations.
Coexisting competing phases are not only often sensitive to static tuning parameters such as pressure, magnetic fields, and chemical doping --- where suppressing or enhancing one phase can significantly modulate the others \cite{Gerber2015Three-dimensionalFields, Kim2018UniaxialSuperconductor} --- but also remarkably susceptible to ultrafast manipulation through light pulses \cite{Jang2022Characterization6.67, Wandel2022EnhancedSuperconductor, Fausti2011Light-inducedCuprate}.
As a prominent example, charge density waves (CDWs) \cite{Gruner2018DensitySolids}, which often compete with other proximate phases, are particularly responsive to optical excitation \cite{Eichberger2010SnapshotsWaves, Lee2012PhaseNickelate, Huber2014CoherentTransition, Singer2016PhotoinducedAmplitude, Vogelgesang2018PhaseDiffraction,Frigge2017OpticallyLimit, Zong2018UltrafastWave, Zong2019EvidenceTransition, Kogar2020Light-inducedLaTe3}. 
In unconventional cuprate superconductors, for instance, CDW coherence can be significantly enhanced by optically quenching superconductivity \cite{Jang2022Characterization6.67, Wandel2022EnhancedSuperconductor}. 
Conversely, phonon-assisted melting of CDWs can strengthen coherent interlayer Josephson coupling, thereby enhancing superconductivity \cite{Fausti2011Light-inducedCuprate}.
These observations suggest the potential for both enhancement and reduction of orders by exploiting optically controllable competing phases.

\begin{figure}[t]
\includegraphics[width=3.375in]{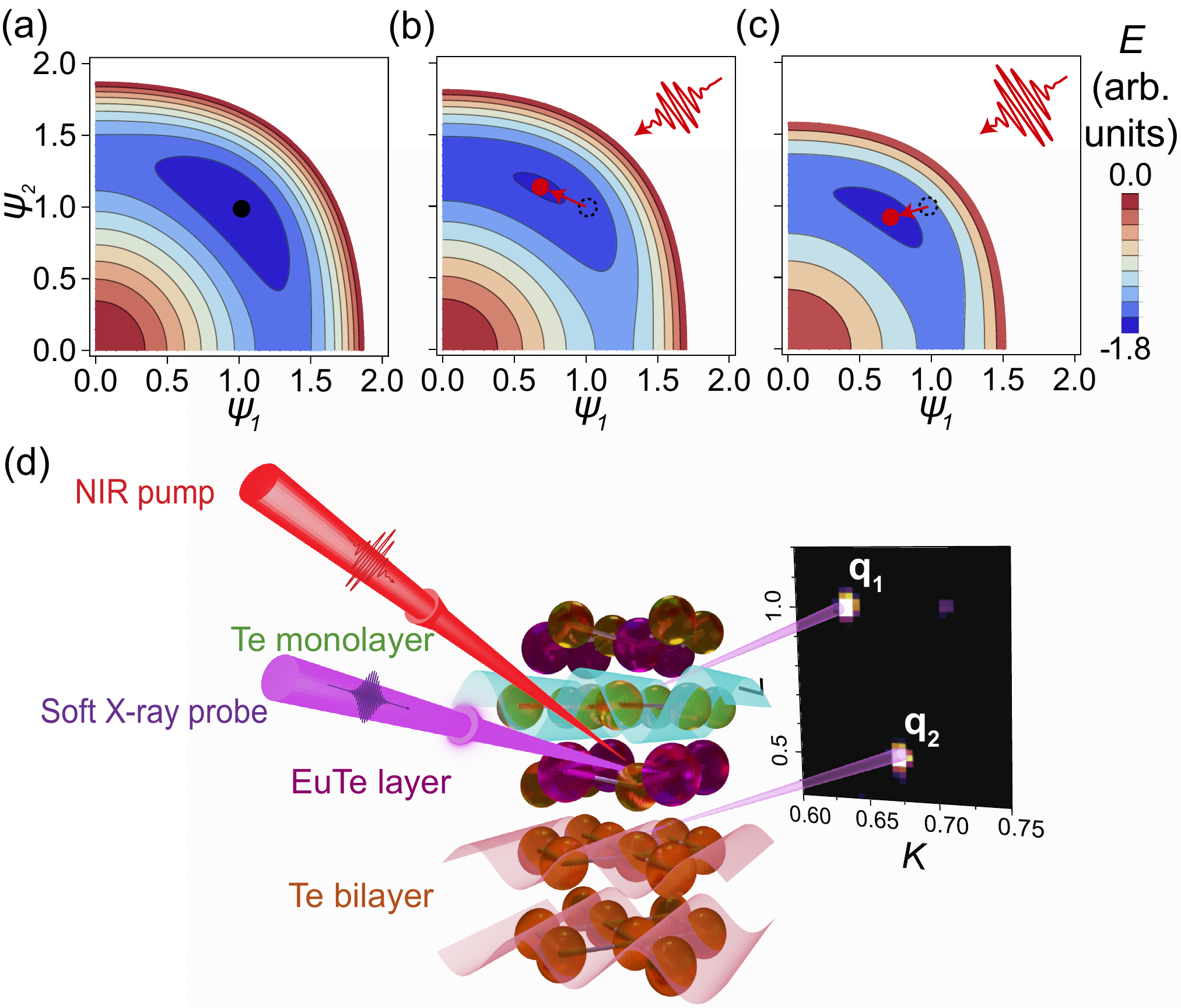}
\caption{Phase competition and experimental setup. (a) Contour plots of equilibrium potential energy surfaces (PES) composed of two competing order parameters $\psi_1$ and $\psi_2$. The black circle marks the static global minimum. (b),(c) Contour plots of excited PES upon a moderate pump and an intense pump, respectively. The red circle marks the excited global minimum. (d) Schematic of the time-resolved X-ray diffraction (tr-XRD) setup with a near-infrared (NIR) pump centered at 800 nm. Eu and Te atoms are colored in purple and yellow, respectively. CDWs residing in the Te monolayers and bilayers are highlighted by blue and red waves, respectively, with their generated diffraction peaks labeled in the static reciprocal space mapping image of the (2 $K$ $L$) plane.}
\label{Fig1}
\end{figure}

Compared to phases of different origins, the coexistence of multiple charge orders, all of which engage electrons near the Fermi level and hence exhibit intrinsic mutual competition, can yield equally rich phenomena \cite{Ning2024DynamicalSuperconductor}. 
In systems with competing CDWs, the amplitude of a more resilient CDW can be enhanced at the expense of a more readily suppressed counterpart, whereas in the absence of phase competition, light pulses tend to restore higher symmetry, thereby reducing the CDW order amplitude. 
This behavior can be rationalized through a phenomenological Ginzburg-Landau theory, where the potential energy $E$ can be expressed as a function of two competing CDW order parameters $\psi_1$ and $\psi_2$ with nonzero values in equilibrium [Fig.~\ref{Fig1}(a)] \cite{Dolgirev2020AmplitudeExperiments, Dolgirev2020Self-similarExperiments, Schafer2010DisentanglementSpectroscopy}.
If $\psi_2$ exhibits significantly lower photo-susceptibility than $\psi_1$, its amplitude can be enhanced upon moderate pump excitation, while $\psi_1$ displays a conventional decrease [Fig.~\ref{Fig1}(b)].
On the other hand, intense excitation that strongly quenches $\psi_1$ and $\psi_2$ to a similar extent leads to a reduction in their amplitudes, though the degree of melting can be different depending on their photo-susceptibilities [Fig.~\ref{Fig1}(c)].
The dynamic evolution of the free-energy landscape, controllable by tailored optical excitation, may thus lead to nonmonotonic, bidirectional order parameter dynamics exhibiting both enhancement and reduction at different excitation intensities --- a phenomenon yet to be realized.

To explore such possibilities, we investigate EuTe$_4$, a newly discovered layered compound hosting multiple incommensurate CDWs \cite{Wu2019LayeredSheets, Lv2022UnconventionalWave, Zhang2022Angle-resolved4, Pathak2022Orbital-/math, Rathore2023NonlocalEuTe4, Zhang2023ThermalEuTe4, Liu2024Room-temperatureWave, Xiao2024HiddenEuTe4, Rathore2023EvolutionEuTe4, Lv2025LargeWaves}.
EuTe$_4$ consists of alternating Te monolayer and bilayer square-net sheets separated by insulating EuTe spacers along the $c$-axis [Fig.~\ref{Fig1}(d)].
Recent photoemission spectroscopy and diffraction measurements have confirmed the existence of two incommensurate CDWs with wavevector $\mathbf{q}_1=$ (0 0.644 0) in the Te monolayer and $\mathbf{q}_2=$ (0 0.678 0.5) in the Te bilayer \cite{Lv2024CoexistenceSemiconductor,Lv2025LargeWaves} [Fig.~\ref{Fig1}(d)]. 
Both CDWs persist beyond 400~K with wavevectors robust against temperature variation \cite{Lv2025LargeWaves}.
Furthermore, these incommensurate wavevectors satisfy the so-called ``joint commensuration'' \cite{Bruinsma1980Phase-lockedNbSe3, FWBoswell1983Charge-densityTaTe4, Emery1979LockingNbSe3}, i.e., the addition of the incommensurate wavevectors equal to a commensurate number: $\mathbf{q_1}+2\mathbf{q_2}=\mathbf{b^*}+\mathbf{c^*}$, where $\mathbf{b^*}$ and $\mathbf{c^*}$ are reciprocal lattice units. 
This joint commensuration relation gives rise to the first observation of moir\'{e} CDW superstructure \cite{Lv2025LargeWaves}.
More importantly, compared to the $\mathbf{q}_1$ CDW, the $\mathbf{q}_2$ CDW exhibits a larger CDW gap and stronger robustness against perturbations due to additional intrabilayer interactions \cite{Lv2024CoexistenceSemiconductor}.
Their different photo-susceptibilities thus provide a playground to search for phase-competition-mediated nonmonotonic dynamical behavior.

Despite these recent advancements, the mutual relationship between the two orders remains unclear.
To address this, a structural probe that is directly sensitive to the CDW order parameter behaviors under perturbation is highly desired, with time-resolved X-ray diffraction (tr-XRD) being one of the most effective tools for this purpose. 
We thus performed tr-XRD measurements in reflection geometry to directly track the temporal evolution of diffraction peaks arising from both orders upon a near-infrared excitation centered around 800~nm under room temperature on the cooling branch  [Fig.~\ref{Fig1}(d)] \cite{Jang2020Time-resolvedLaser,Lv2022UnconventionalWave} (See Supplemental Material \footnote{\label{note1} See Supplemental Material for additional details on the experimental setups, theoretical background, and discussions. It includes 8 sections: (i) Experimental details, (ii) Temporal evolution of the CDW peak profile, (iii) Structural Bragg peak dynamics, (iv) Excluding coherent oscillation as an origin of the dip-peak-hump feature, (v) Fitting procedure of the $\mathbf{q_1}$ and $\mathbf{q_2}$ peak dynamics, (vi) Light-induced quenching and recovery dynamics of the $\mathbf{q_2}$ peak, (vii) Ginzburg-Landau theory for jointly commensurate CDWs, (viii) Absence of non-volatility in our experimental geometry}). 
Our results unambiguously unveil the competition between the two incommensurate CDWs and demonstrate the tunability of the bilayer CDW amplitude as a function of time and photo-excitation intensity.

\begin{figure}[t]
\includegraphics[width=3.375in]{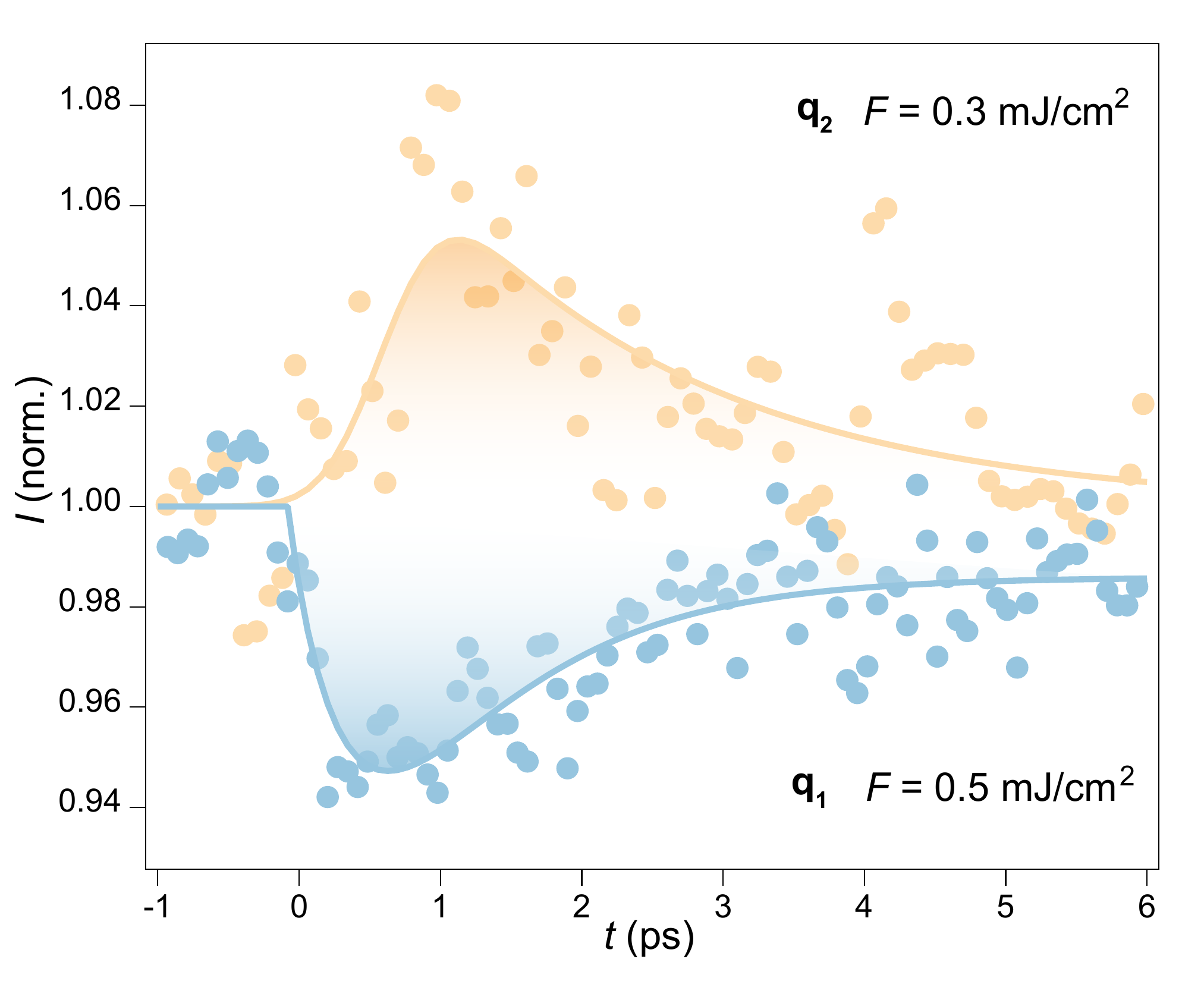}
\caption{Temporal evolution of the $\mathbf{q_1}$ and $\mathbf{q_2}$ peak intensities $I$ normalized to their equilibrium values. Plots are taken with the pump fluence $F$ = 0.5 mJ/cm$^2$ and 0.3 mJ/cm$^2$, respectively, which are the lowest pump fluence employed in this work. Solid curves are fits to equations in \hyperref[note1]{[35]}.}
\label{Fig2}
\end{figure}

\begin{figure*}[t]
\includegraphics[width=6.75in]{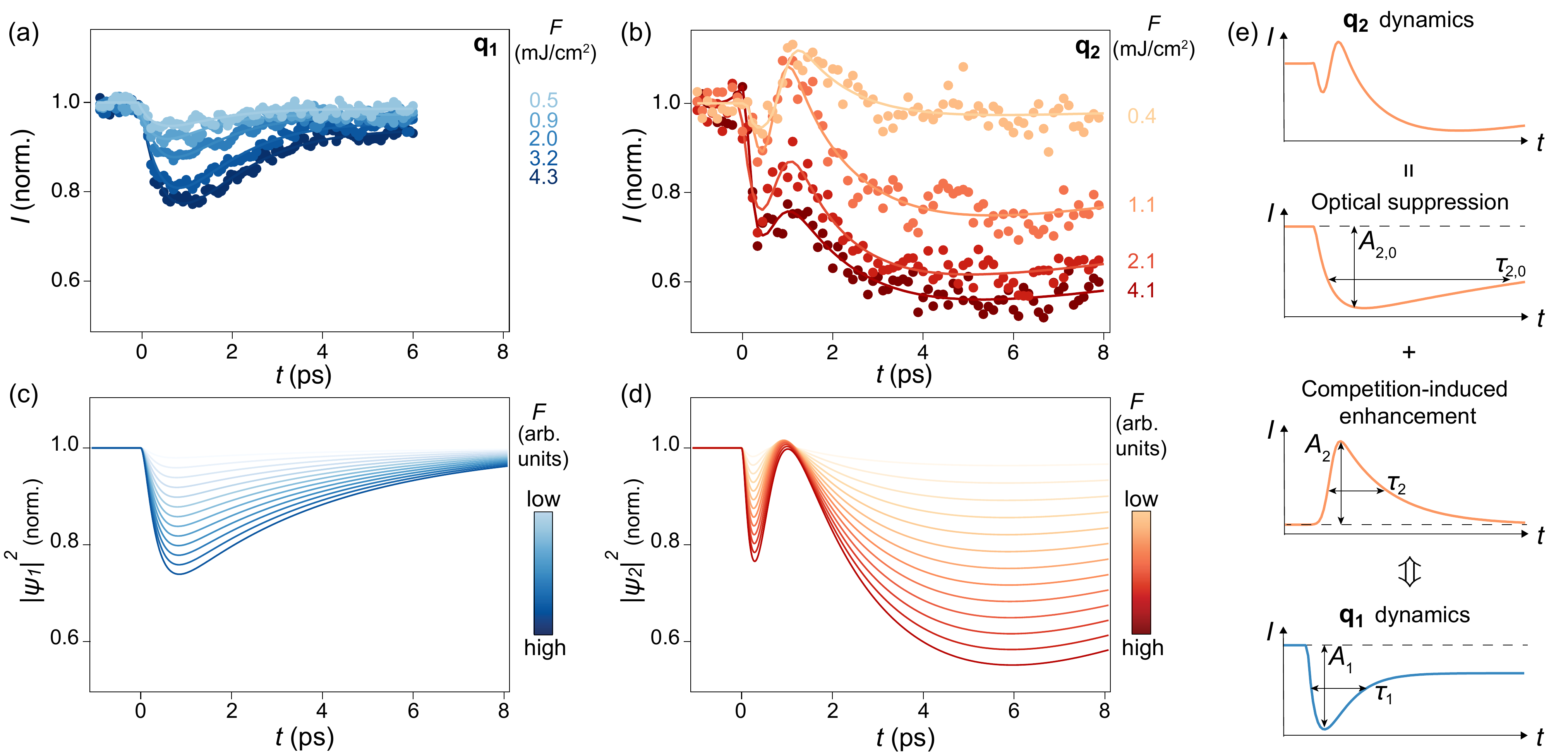}
\caption{Pump fluence ($F$) dependent CDW dynamics. (a),(b), Temporal evolution of the intensity $I$ acquired from the tr-XRD measurements normalized to their equilibrium values for selected $F$ of the $\mathbf{q_1}$ and $\mathbf{q_2}$ peaks, respectively. Solid curves are fits to equations in \hyperref[note1]{[35]}. (c),(d), Temporal evolution of the normalized order parameter amplitude square $|\psi|^2$ obtained by TDGL simulations for selected $F$ of the CDW order parameters $\psi_1$ and $\psi_2$. Fluence ranges are identical for both order parameters. (e) Schematics of the decomposition of the $\mathbf{q_2}$ peak dynamics. The $\mathbf{q_2}$ peak dynamics can be considered as a summation of a photo-induced suppression and a phase-competition-induced enhancement that anticorrelates with the $\mathbf{q_1}$ peak suppression. Different fitting parameters defined in \hyperref[note1]{[35]} are shown.}
\label{Fig3}
\end{figure*}

We first investigate the temporal evolution of the integrated intensity of the CDW diffraction peak at around (0 -0.644 1), which corresponds to the CDW with wavevector $\mathbf{q_1}$, following a pump fluence of $F=0.5$~mJ/cm$^2$ (Fig.~\ref{Fig2}). 
As expected from a conventional light-induced suppression of CDW, the intensity decreases by $\sim$6$\%$ at $\sim$0.6 ps before relaxing to a quasi-equilibrium value over $\sim$2 ps \cite{Eichberger2010SnapshotsWaves, Huber2014CoherentTransition, Zong2019EvidenceTransition}.
On the other hand, the peak at around (0 0.678 1.5), corresponding to the CDW with wavevector $\mathbf{q_2}$, exhibits a behavior distinct from conventional light-induced suppression.
With a pump fluence of $F = 0.3$~mJ/cm$^2$, we observe a transient increase of approximately $\sim$6$\%$ within $\sim$1.2 ps, followed by a recovery over $\sim$2 ps back to the equilibrium value.
While the initial decay of the $\mathbf{q_1}$ peak is slightly faster than the rise of the $\mathbf{q_2}$ peak, both share a similar recovery time.
This overall synchronized dynamics accompanied by a disparity in initial timescale, closely aligns with previous observations in YBa$_2$Cu$_3$O$_{6+x}$ and LaTe$_3$ \cite{Jang2022Characterization6.67, Wandel2022EnhancedSuperconductor, Kogar2020Light-inducedLaTe3}, which is attributed to the competitive interplay between two coexisting phases.
We note that both the $\mathbf{q_1}$ and $\mathbf{q_2}$ peaks exhibit no measurable change in peak position or width along the in-plane $K$ and out-of-plane $L$ directions, indicating that the measured intensity change purely arises from the CDW order parameter amplitude change or phase fluctuations \footnotemark[1].
Furthermore, such melting and recovery dynamics of CDWs are highly nonthermal, as structural Bragg peaks do not exhibit quick dynamics stemming from laser-induced heating over the same timescale  \footnotemark[1].

To more comprehensively understand the correlation between the two peaks, we measure their intensity dynamics upon various pump fluences $F$.
The $\mathbf{q_1}$ peak displays a consistent photo-induced melting, which increases with $F$, and a subsequent partial recovery to a quasi-equilibrium [Fig.~\ref{Fig3}(a)].
On the contrary, a dramatic nonmonotonic behavior emerges in the dynamics of the $\mathbf{q_2}$ peak.
At $F=0.4$ mJ/cm$^2$, prior to the enhancement up to $\sim$10\% and the following recovery [Fig.~\ref{Fig3}(b)], a faint dip occurs around $t=0.3$ ps, which is not observed in the lower fluence time trace within the signal-to-noise ratio [Fig.~\ref{Fig2}(a)].
As $F$ further increases, this nonmonotonic behavior becomes more pronounced.
The $\mathbf{q_2}$ peak intensity first decreases within 0.3 ps and the amplitude of this dip increases with $F$.
Subsequently, the intensity increases and reaches a maximum around $t=1.2$ ps.
Ensuing the peak value, the intensity then decreases again within 5 ps, which later recovers to the equilibrium value over tens of picoseconds.
Such a ``dip-peak-hump'' lineshape does not arise from overdamped coherent phonons, as coherent phonon oscillations of the diffraction intensity observed in this material last up to 8 ps \footnotemark[1].

We note that the dynamics of the $\mathbf{q_2}$ peak can be phenomenologically decomposed into a negative decay and recovery and a positive rise and recovery [Fig.~\ref{Fig3}(e)].
The former can be attributed to the intrinsic photo-quenching of the corresponding order parameter, while the latter seems closely anti-correlated with the $\mathbf{q_1}$ peak dynamics across a broad fluence range, reminiscent of the light-induced enhancement of CDW amplitude arising from the suppression of a competing phase \cite{Jang2022Characterization6.67,Wandel2022EnhancedSuperconductor, Kogar2020Light-inducedLaTe3}.
We thus tentatively propose a picture involving both light-induced suppression and phase-competition-induced enhancement of order parameters to explain the nontrivial temporal evolution of the $\mathbf{q_1}$ and $\mathbf{q_2}$ peaks:
Immediately after light excitation, the CDW order parameter amplitudes contributing to both the $\mathbf{q_1}$ and $\mathbf{q_2}$ peaks are quenched, with the latter decaying faster, as indicated by the experimental data. 
Due to their mutual competition, suppression of the $\mathbf{q_1}$ peak tends to enhance the $\mathbf{q_2}$ peak, counteracting its light-induced melting and resulting in the emergence of a local ``dip'' feature in the $\mathbf{q_2}$ peak time trace.
Subsequently, the $\mathbf{q_2}$ peak intensity reaches its local maximum at $\sim$1 ps, yielding a delayed ``peak'' feature.
Meanwhile, the intensity of the $\mathbf{q_1}$ peak continues to decrease for up to $\sim$1 ps.
Then, as the $\mathbf{q_1}$ peak intensity recovers, competition starts to suppress the $\mathbf{q_2}$ peak intensity, yielding its reduction until $\sim6$ ps when the $\mathbf{q_1}$ peak reaches quasi-equilibrium.
Eventually, before the $\mathbf{q_2}$ peak intensity recovers to its equilibrium value on a much longer timescale, a second minimum resembling a ``hump'' is generated.
Here, the asymmetric dynamics between the two peaks stem from the larger optical susceptibility of the $\mathbf{q_1}$ CDW compared to the $\mathbf{q_2}$ CDW, reducing the prominence of phase competition effects in the $\mathbf{q_1}$ peak intensity dynamics. The lower photo-susceptibility of the $\mathbf{q_2}$ CDW is consistent with the previous joint ultrafast photoemission and density functional theory study, as the Te bilayer CDW is further stabilized by intrabilayer interactions \cite{Lv2024CoexistenceSemiconductor}.

\begin{figure}[t]
\includegraphics[width=3.375in]{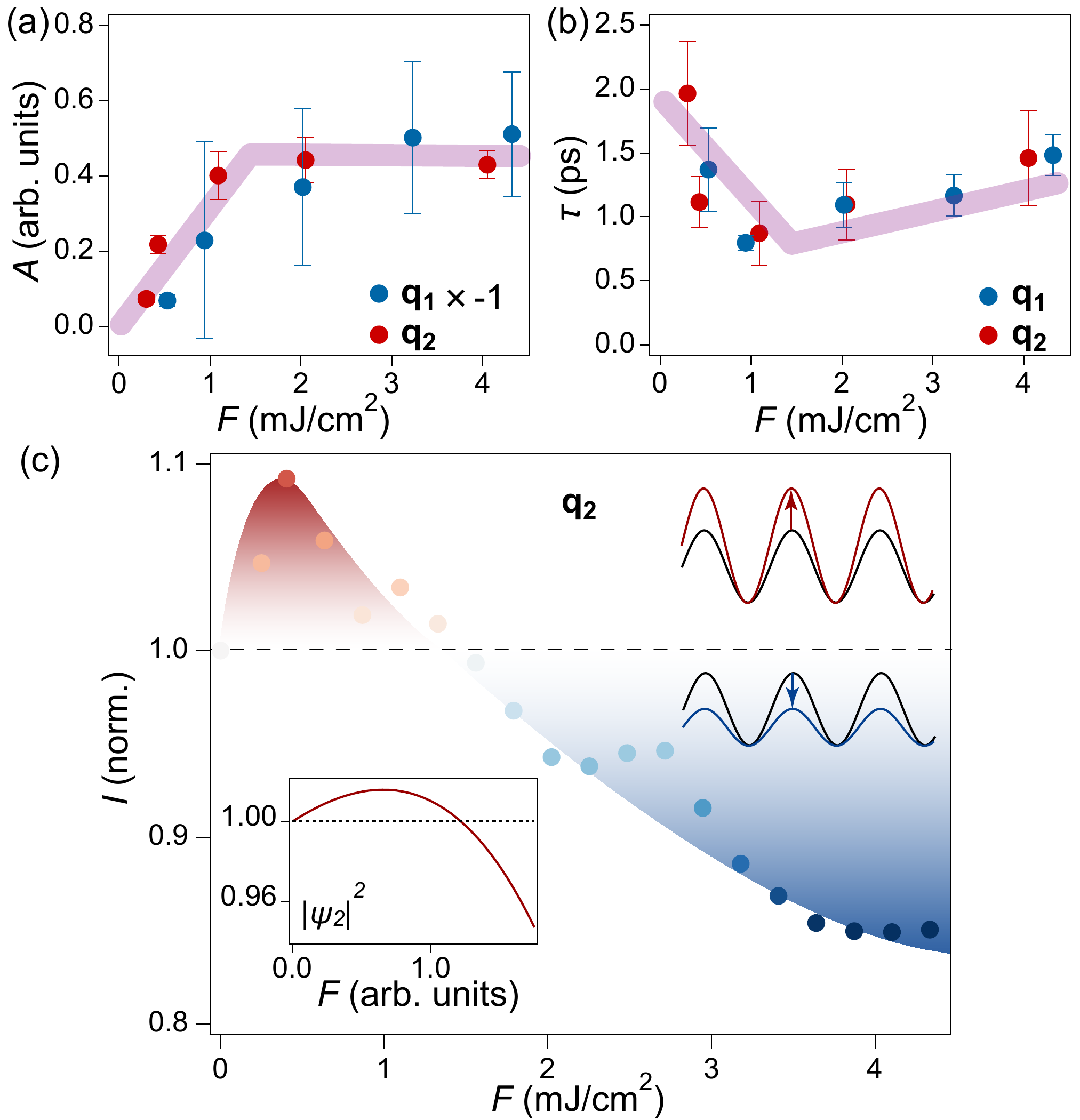}
\caption{(a) Pump fluence ($F$) dependence of the fitted magnitudes $A_1$ and $A_2$ of the $\mathbf{q_1}$ and $\mathbf{q_2}$ peaks. The sign of $A_1$ is flipped for direct comparison with $A_2$. (b) $F$ dependence of the fitted decay times $\tau_1$ and $\tau_2$ of the $\mathbf{q_1}$ and $\mathbf{q_2}$ peaks. Broad solid lines are guides to eyes. (c) $F$ dependence of the transient normalized intensity of the $\mathbf{q_2}$ peak at $t=1.1$ ps. Left inset shows the simulated fluence dependence of the CDW order amplitude square at a fixed time delay where the ``peak'' feature is maximal. Red and blue shadings highlight the light-enhanced and reduced regions, respectively, as pictorially shown by the right inset.}
\label{Fig4}
\end{figure}

If this proposed picture holds true, the amplitude and recovery timescale of the positive enhancement of $\mathbf{q_2}$ should mirror those of $\mathbf{q_1}$.
To verify this scenario, we fit the intensity time trace of the $\mathbf{q_2}$ peak to a superposition of negative and positive exponentials to describe the suppression and enhancement, respectively, and obtain the CDW enhancement magnitude, $A_{2}$, and the timescale, $\tau_{2}$, from the positive exponential term  [Fig.~\ref{Fig3}(e)] \footnotemark[1].
We also fit the intensity time trace of the $\mathbf{q_1}$ peak using a single negative exponential to extract the CDW suppression magnitude, $A_{1}$, and the timescale, $\tau_{1}$ [Fig.~\ref{Fig3}(e)].
We then compare the fitted values of ($A_{1}$, $\tau_{1}$) with ($A_{2}$, $\tau_{2}$) across various $F$ values to scrutinize their correlation.
Remarkably, $F$-dependencies of both quantities for both peaks exhibit a close resemblance, with $A$ saturating and $\tau$ minimizing at around $F$= 1.5 mJ/cm$^2$ [Figs.~\ref{Fig4}(a) and \ref{Fig4}(b)]. 
The amplitude inflection may arise from absorption saturation, as neither of the CDW orders is completely suppressed at this fluence.
The relatively fast timescale of $\tau$ associated with its moderate fluence dependence also suggests that it mainly reflects the relaxation of charge order amplitude, as phase fluctuations typically recover at longer timescales and exhibit more significant increase as the pump fluence increases \cite{Zong2019EvidenceTransition,Lee2012PhaseNickelate}.
Altogether, such synchronized and proportional change of both peak intensities corroborates the proposed mechanism, benchmarking the strong coupling and competition between the two orders.
In comparison, the intrinsic CDW suppression extracted from the negative exponential of the $\mathbf{q_2}$ peak does not show strong correlation with those of the $\mathbf{q_1}$ peak \footnotemark[1].

\paragraph{}

To rationalize the proposed picture within a theoretical framework, we first develop a phenomenological Ginzburg-Landau theory involving complex order parameters $\psi_1$ and $\psi_2$, corresponding to the orders resulting in the $\mathbf{q_1}$ and $\mathbf{q_2}$ peaks, respectively. 
Based on symmetry analysis, the free energy functional expanded up to the fourth order in $\psi$'s can be expressed as (see details in Supplemental Material \footnotemark[1]):
\begin{equation}\label{eqLG}
\begin{split}
    E=&-\frac{1}{2}a_1|\psi_1|^2+\frac{1}{4}b_1|\psi_1|^4-\frac{1}{2}a_2|\psi_2|^2+\frac{1}{4}b_2|\psi_2|^4 \\
    &+c(\psi_1\psi_2\psi_2+\psi_1^*\psi_2^*\psi_2^*)+d\psi_1\psi_1^*\psi_2\psi_2^*\\
\end{split}
\end{equation}
where $a_1$, $b_1$, $a_2$, $b_2$, $c$, and $d$ are real coefficients.
Here, the first four terms characterize the Mexican hat potential for each order parameter without considering their coupling. 
The cubic term arises from the special wavevector constraints of joint commensuration, with the inclusion of the complex conjugate term to preserve inversion symmetry.
The remaining quartic term is the coupling that is always allowed independent of the wavevectors.
We set $a_i>0$ and $b_i>0$ so that the potential features a double-well shape for all $\psi$'s in the absence of coupling terms ($c=0, d=0$), which gives rise to the simplest description of a second-order phase transition. 
We also set a positive $d>0$, imposing a competing relationship between two orders.
Based on the initial suppression of both peaks, we assume that optical excitation quenches both $a_1$ and $a_2$ and thus modifies the Landau energy functional, stimulating the order parameters to seek a new minimum. 
We also note that while the other terms are symmetric for both order parameters, the cubic term introduces an asymmetry between $\psi_1$ and $\psi_2$, making the former more susceptible to light excitation and further contributing to the asymmetric evolution.
The temporal evolution of $\psi_1$ and $\psi_2$ can then be derived through the typical treatment of solving a time-dependent Ginzburg-Landau (TDGL) theory, $\partial_t\psi_i=-\gamma_i \partial E/\partial \psi_i^*$ ($i$=1,2), where $\gamma_1$ and $\gamma_2$ are phenomenological damping parameters.
Our simulated order parameter dynamics reveal a nonmonotonic behavior of $\psi_2$ and partial melting of $\psi_1$ [Figs.~\ref{Fig3}c,d], in close agreement with the experimental observations [Figs.~\ref{Fig3}a,b]. 
With a reasonable selection of parameters, not only the relative melting levels but also the relative timescales are qualitatively reproduced.
These results further confirm that the abnormal behavior of the $\mathbf{q_2}$ peak emerges from the phase competition.

While our minimal theory well captures the essential physics, a quantitative mismatch of the ``peak'' value around 1 ps with the experimental data remains.
We find that a more quantitative match can be achieved by further assuming pump-fluence-dependent CDW melting and recovery times and the coupling $d$ decreasing with pump fluence as photoexcitation weakens the interlayer coupling through enhanced electron screening (Supplemental Material \footnotemark[1]).
Additional ingredients including extensions beyond the mean-field limit, time-dependent damping \cite{Huber2014CoherentTransition}, and anharmonic phonon coupling \cite{Hase2002DynamicsPhotoexcitation} could be further incorporated in the simulations for improved quantitative agreement.

Our joint experimental and theoretical study thus showcases the competition between the two CDWs in EuTe$_4$. 
As the $\mathbf{q_1}$ and $\mathbf{q_2}$ CDWs reside in different Te layers and consume electronic states from distinct Te $p$-orbitals, the usual argument for CDW competition due to the consumption of density of states near Fermi level does not seem to apply here. 
Nonetheless, prior studies have shown that the two orders can interact, forming interlayer excitons and enabling charge transfer between the Te monolayer and bilayer \cite{Lv2024CoexistenceSemiconductor,Lv2025LargeWaves}.
While a detailed microscopic theory is still needed to fully elucidate this competition, the intrinsic interlayer coupling between the two CDWs may serve as the underlying driver, resembling cases where competing CDWs emerge in the same layer. 

The dynamic enhancement and suppression of the $\psi_2$ amplitude suggest that it can be controlled by adjusting the total photo-excitation intensity.
We demonstrate this by tracking the $\mathbf{q_2}$ peak intensity as a function of the pump fluence $F$ at $t=1.1$ ps, where the largest enhancement is observed [Fig.~\ref{Fig4}(c)].
The intensity initially increases with $F$ up to 0.5 mJ/cm$^2$, and then decreases, returning to unity around $F=1.5$ mJ/cm$^2$.
Larger $F$ further reduces the order parameter amplitude below the equilibrium value, with a potential saturation above 4 mJ/cm$^2$.
We also track the temporal evolution of $|\psi_2|^2$ simulated using TDGL theory, and this nonmonotonic dependence on $F$ can be reproduced [Fig.~\ref{Fig4}(c), left inset].
A closer examination reveals that the amplitude saturation [Fig.~\ref{Fig4}(a)], upturn in decay time  [Fig.~\ref{Fig4}(b)], and the crossover from enhancement to reduction [Fig.~\ref{Fig4}(c)] all occur at around $F=1.5$ mJ/cm$^2$, whose underlying mechanism requires further investigation.

In conclusion, our tr-XRD results uncover anomalous nonmonotonic dynamics of the bilayer CDW in EuTe$_4$ mediated by light-induced melting and phase competition. 
We demonstrate versatile accessibility of both order amplitude enhancement and reduction as functions of time or excitation intensity.
These findings provide a smoking gun clarification of the relationship between different CDWs phases in this novel compound, which can be extensively applied to a plethora of exotic materials that host multiple CDW orders, such as CsV$_3$Sb$_5$ \cite{Ning2024DynamicalSuperconductor}, BaNi$_2$As$_2$ \cite{Lee2019UnconventionalBaNi1xCox2As2, Lee2021MultipleBa1xSrxNi2As2}, and NbSe$_2$ \cite{Gye2019Topological/math, Guster2019Coexistence2}. 
Our demonstrated ability to selectively quench one CDW amplitude may provide a route to engineer non-volatile states observed in EuTe$_4$ \cite{Liu2024Room-temperatureWave,Lv2022UnconventionalWave} by first fully suppressing and then phase-inverting a single CDW using intense pulses \footnotemark[1].
Moreover, probably due to the separated spatial location, EuTe$_4$ constitutes an unconventional example compared to the other CDWs where the enhancement and reduction of the order parameter occur without coherence and wavevector alterations.
The easily exfoliable structure and tunable CDW amplitudes above and below the equilibrium value also make the targeted material promising for integration into quantum electronic devices \cite{Ogawa2002Charge-densityMemory, Liu2016ATemperature, Khitun2017Two-DimensionalDevices, Geremew2019Bias-VoltageDevices, Wang2019WritingPulses, Mihailovic2019TheApplications}.

\section*{Acknowledgments}
The authors thank Tianchuang Luo, Batyr Ilyas, Bryan Fichera, Mingu Kang, Pavel Volkov, and Zhuquan Zhang for fruitful discussions. The work at MIT was supported by the U.S. Department of Energy, the BES DMSE (data collection and analysis) and the Gordon and Betty Moore Foundation's EPiQS Initiative grant GBMF9459 (manuscript writing). Research conducted at the Center for High-Energy X-ray Science (CHEXS) is supported by the National Science Foundation (BIO, ENG and MPS Directorates) under award DMR-2342336. H.J. acknowledges support by the National Research Foundation grant funded by the Korea government (MSIT) (Grant No. RS-2022-NR068223). The work conducted at the Center for High Energy X-ray Sciences (CHEXS) at CHESS is supported by the National Science Foundation under award DMR-1829070. The tr-XRD experiments were performed at the SSS-RSXS end station (Proposal No. 2024-1st-SSS-008) of the PAL-XFEL funded by the Korea government (MSIT). N.L.W acknowledges support from National Natural Science Foundation of China (Grant No. 12488201). N.L.W and D.W are supported by National Key Research and Development Program of China (2024YFA1408700).

\end{document}